# Silicon photonic filters based on cascaded Sagnac loop resonators

Jiayang Wu,[1] Tania Moein,[1] Xingyuan Xu,[1] and David J. Moss[1,a]

[1]Centre for Micro-Photonics, Swinburne University of Technology, Hawthorn, Victoria 3122, Australia

We demonstrate advanced integrated photonic filters in silicon-on-insulator (SOI) nanowires implemented by cascaded Sagnac loop reflector (CSLR) resonators. We investigate mode splitting in these standing-wave (SW) resonators and demonstrate its use for engineering the spectral profile of on-chip photonic filters. By changing the reflectivity of the Sagnac loop reflectors (SLRs) and the phase shifts along the connecting waveguides, we tailor mode splitting in the CSLR resonators to achieve a wide range of filter shapes for diverse applications including enhanced light trapping, flat-top filtering, Q factor enhancement, and signal reshaping. We present the theoretical designs and compare the CSLR resonators with three, four, and eight SLRs fabricated in SOI. We achieve versatile filter shapes in the measured transmission spectra via diverse mode splitting that agree well with theory. This work confirms the effectiveness of using CSLR resonators as integrated multi-functional SW filters for flexible spectral engineering.

## I. INTRODUCTION

Integrated photonic resonators have attracted great interest for their applications in signal modulation, buffering, switching, and processing in optical communications systems.[1,2] They have been enabled by advanced micro/nano fabrication technologies and offer compact footprints, mass producibility, scalability, and versatile filtering properties. Mode splitting induced by coherent optical mode interference in coupled resonant cavities is a key phenomenon in photonic resonators that can lead to powerful and versatile filtering functions such as optical analogous to electromagnetically-induced-transparency (EIT), Fano resonances, Autler-Townes splitting, and dark states.[3-5] It is similar in principle to atomic resonance splitting caused by quantum interference between excitation pathways in multi-level atomic systems.[6] One great advantage of this effect is that it can break the dependence between quality factor, free spectral range (FSR), and physical cavity length.[7-9] Moreover, the resulting group delay response and mode interaction are useful for enhancing light-material interaction and dispersion engineering in nonlinear optics.[10-14]

Photonic resonators can be classified into two categories − travelling-wave (TW) resonators, exemplified by ring resonators, and standing-wave (SW) resonators represented by photonic crystal cavities, distributed feedback cavities, and Fabry-Pero (FP) cavities.[3] The majority of work on mode splitting in photonic resonators has been based on TW resonators[15-19] although some recent work has investigated device structures consisting of both TW and SW resonators.[20-22] Since TW resonators are almost twice as long as their SW counterparts for the same FSR,[23] SW resonators tend to attract more interest in terms of device footprint. In addition, unlike the nearly uniform field distribution in TW resonators, the spatially dependent field distribution in SW resonators is useful for the efficient excitation of laser emission and nonlinear optical effects.[24-28]

---

[a]Author to whom correspondence should be addressed. Electronic mail: dmoss@swin.edu.au

In this paper, we demonstrate advanced integrated photonic filters in silicon-on-insulator (SOI) nanowires by exploiting mode splitting in SW resonators formed by cascaded Sagnac loop reflectors (SLRs), which we term cascaded SLR (CSLR) resonators. We show that this is a powerful approach for engineering the spectral profile of on-chip filters. The concept of CSLR resonators was proposed in Refs. 23 and 17, and experimental demonstrations of CSLR resonators with two SLRs were reported in Refs. 23 and 29. For CSLR resonators with more than two SLRs, mode splitting occurs due to coherent interference between the FP cavities formed by the different SLRs. Here, we provide a detailed theory of mode splitting in CSLR resonators for spectral engineering, and experimentally demonstrate CSLR resonators with up to eight SLRs in SOI based nanowires. By changing the reflectivity of the SLRs and the phase shifts along the connecting waveguides, mode splitting in the CSLR resonators can be tailored for diverse applications such as enhanced light trapping, flat-top filtering, Q factor enhancement, and signal reshaping. We achieve high performance and versatile filter shapes that correspond to diverse mode splitting conditions. The experimental results confirm the effectiveness of our approach towards realizing integrated multi-functional SW filters for flexible spectral engineering.

## II. DEVICE CONFIGURATION AND OPERATION PRINCIPLE

Figure 1 illustrates the schematic configuration of the integrated CSLR resonator. It consists of $N$ SLRs (SLR$_1$, SLR$_2$, …, SLR$_N$) formed by a self-coupled nanowire waveguide loop. In the CSLR resonator, each SLR performs as a reflection/transmission element and contributes to the overall transmission spectra from port IN to port OUT in Fig. 1. Therefore, the cascaded SLRs with a periodic loop lattice show similar transmission characteristics to that of photonic crystals.[17] The two adjacent SLRs together with the connecting waveguide form a FP cavity, thereby $N$ cascaded SLRs can also be regarded as $N$-1 cascaded FP cavities (FPC$_1$, FPC$_2$, …, FPC$_{N-1}$), similar to Bragg gratings.[30, 31] To study the CSLR resonator based on the scattering matrix method,[32-33] we define the waveguide and coupler parameters of the CSLR resonator in Table I, and so the field transmission function from port IN to port OUT can be written as:

$$T_{\text{CSLR}}(N) = \begin{cases} \dfrac{T_{s1} T_{s2} T_1}{1 - R_{s1} R_{s2} T_1^2}, & N = 2 \\ \dfrac{T_{\text{CLSR}}(N-1) T_{sN} T_{N-1}}{1 - R_{\text{CSLR-}}(N-1) R_{sN} T_{N-1}^2}, & N > 2 \end{cases} \quad (1)$$

**TABLE I. Definitions of waveguide and coupler parameters of the CSLR resonator**

| Waveguide | Length | Transmission factor[a] | Phase shift[b] |
|---|---|---|---|
| waveguide connecting $SLR_i$ to $SLR_{i+1}$ ($i$ = 1, 2, .., $N$-1) | $L_i$ | $a_i$ | $\varphi_i$ |
| Sagnac loops in $SLR_i$ ($i$ = 1, 2, .., $N$) | $L_{si}$ | $a_{si}$ | $\varphi_{si}$ |
| Coupler | Coupling length[c] | Field transmission coefficient[d] | Field cross-coupling coefficient[d] |
| couplers in $SLR_i$ ($i$ = 1, 2, .., $N$) | $L_{ci}$ | $t_i$ | $\kappa_i$ |

[a] $a_i = exp(-\alpha L_i /2)$, $a_{si} = exp(-\alpha L_{si} /2)$, $\alpha$ is the power propagation loss factor.
[b] $\varphi_i = 2\pi n_g L_i /\lambda$, $\varphi_{si} = 2\pi n_g L_{si} /\lambda$, $n_g$ is the group index and $\lambda$ is the wavelength.
[c] $L_{ci}$ ($i$ = 1, 2, .., $N$) are the straight coupling lengths shown in Fig.1. They are included in $L_i$.
[d] In our calculation, we assume $t_i^2 + \kappa_i^2 =1$ for lossless coupling in all the directional couplers.

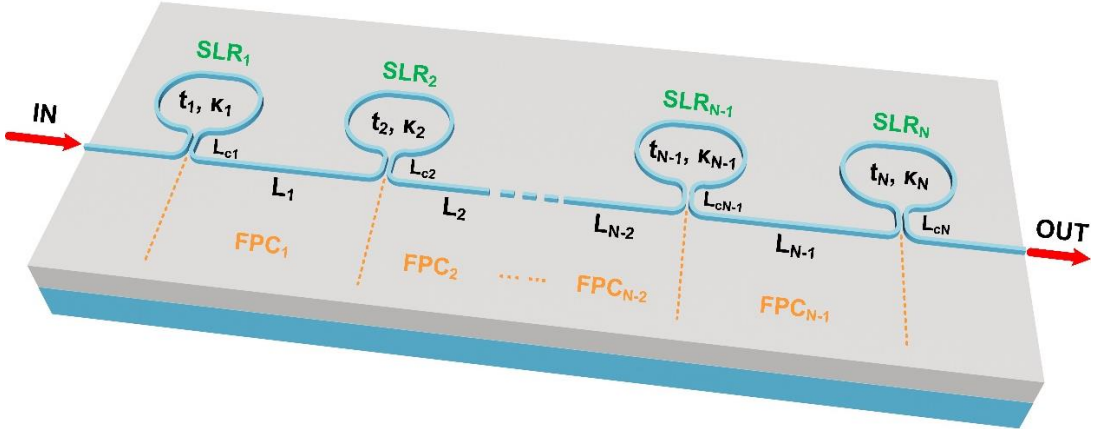

FIG. 1. Schematic configuration of integrated CSLR resonator made up of $N$ cascaded SLRs ($SLR_1$, $SLR_2$, …, $SLR_N$). $FPC_i$ ($i$ = 1, 2, .., $N$-1) are the FP cavities formed by $SLR_i$ and $SLR_{i+1}$, respectively. The definitions of $t_i$ ($i$ = 1, 2, .., $N$), $\kappa_i$ ($i$ = 1, 2, .., $N$), $L_{ci}$ ($i$ = 1, 2, .., $N$), and $L_i$ ($i$ = 1, 2, .., $N$-1) are given in Table I.

In Eq. (1), $T_{si}$ and $R_{si}$ ($i$ = 1, 2, .., $N$) denote the field transmission and reflection functions of $SLR_i$ given by:

$$T_{si} = (t_i^2 - \kappa_i^2)a_{si}e^{-j\varphi_{si}}, \ i = 1, 2, .., N \qquad (2)$$

$$R_{si} = 2jt_i\kappa_i a_{si}e^{-j\varphi_{si}}, \ i = 1, 2, .., N \qquad (3)$$

$T_i$ ($i$ = 1, 2, .., $N$-1) represent the field transmission functions of the waveguide connecting SLR$_i$ to SLR$_{i+1}$, which can be expressed as:

$$T_i = a_i e^{-j\varphi_i}, \quad i = 1, 2, .., N\text{-}1 \tag{4}$$

For the CSLR resonators implemented by SLR$_1$, SLR$_2$, …, and SLR$_i$ ($i$ =1, 2, .., $N$), $T_{\text{CSLR}}(i)$ are the field transmission functions, $R_{\text{CSLR+}}(i)$ and $R_{\text{CSLR-}}(i)$ are the field reflection functions for light input from left and right sides, respectively, which can be given by:

$$R_{\text{CSLR-}}(i) = \begin{cases} \dfrac{R_{s2} + (T_{s2}^2 - R_{s2}^2) T_1^2 R_{s1}}{1 - R_{s1} R_{s2} T_1^2}, & i = 2 \\ \dfrac{R_{si} + (T_{si}^2 - R_{si}^2) T_{i-1}^2 R_{\text{CSLR-}}(i-1)}{1 - R_{\text{CSLR-}}(i-1) R_{si} T_{i-1}^2}, & i > 2 \end{cases} \tag{5}$$

$$R_{\text{CSLR+}}(i) = \begin{cases} \dfrac{R_{s1} + (T_{s1}^2 - R_{s1}^2) T_1^2 R_{s2}}{1 - R_{s1} R_{s2} T_1^2}, & i = 2 \\ \dfrac{R_{\text{CSLR+}}(i-1) + [T_{\text{CLSR}}(i-1)^2 - R_{\text{CSLR+}}(i-1) R_{\text{CSLR-}}(i-1)] T_{i-1}^2 R_{si}}{1 - R_{\text{CSLR-}}(i-1) R_{si} T_{i-1}^2}, & i > 2 \end{cases} \tag{6}$$

In Eqs. (2) and (3), it can be seen that the transmittance and reflectivity of the SLR$_i$ depend on the $t_i$, In terms of practical fabrication, the $t_i$ can be engineered by changing the coupling length $L_{ci}$. The large dynamic range in the transmittance and reflectivity of individual SLRs that can be engineered by changing $t_i$ or $\kappa_i$ makes the CSLR resonator more flexible for spectral engineering as compared with Bragg gratings. On the other hand, according to Eq. (4), the transmission spectra of the CSLR resonators can also be tailored by changing $\varphi_i$ ($i$ = 1, 2, .., $N$-1) - i.e., the phase shifts along the connecting waveguides. The freedom in designing $t_i$ ($i$ =1, 2, .., $N$) and $\varphi_i$ ($i$ = 1, 2, .., $N$-1) is the basis for flexible spectral engineering based on the CSLR resonators, which can lead to versatile applications. In Eqs. (5) and (6), $R_{\text{CSLR+}}(i)$ equals to $R_{\text{CSLR-}}(i)$ only when SLR$_1$, SLR$_2$, …, and SLR$_i$ are identical, i.e., when SLR$_1$, SLR$_2$, …, and SLR$_i$ are not identical, there are nonreciprocal reflections from the CSLR resonators for light input from different directions. These non-reciprocal reflections are induced by different losses within the resonant cavity - the CSLR resonators themselves will still have reciprocal transmission for light input from different directions.

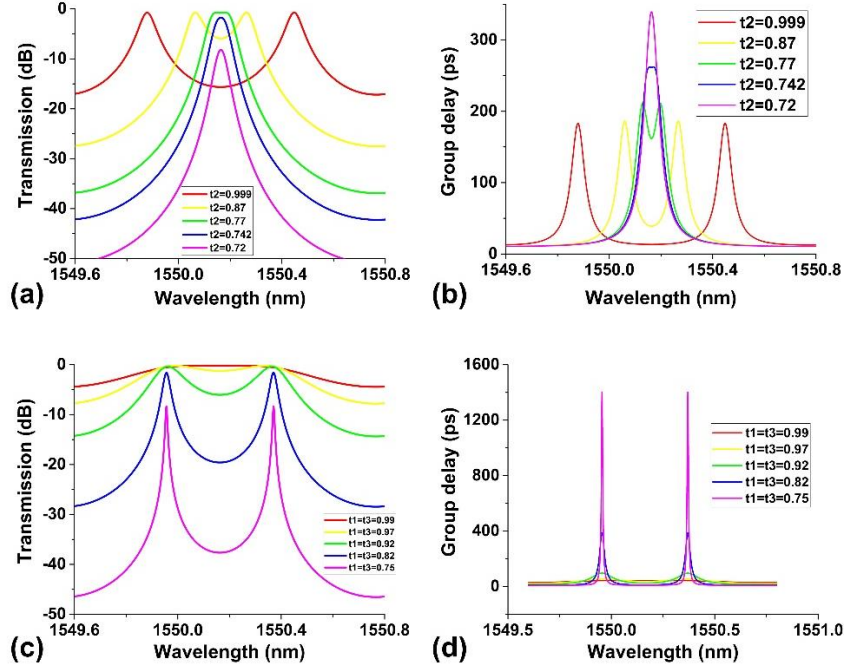

FIG. 2. (a) Calculated power transmission spectra of the CSLR resonator ($N$ = 3) for various $t_2$ when $t_1 = t_3 = 0.87$. (b) Calculated group delay spectra of the CSLR resonator ($N$ = 3) for various $t_2$ when $t_1 = t_3 = 0.87$. (c) Calculated power transmission spectra of the CSLR resonator ($N$ = 3) for various $t_1 = t_3$ when $t_2 = 0.97$. (d) Calculated group delay spectra of the CSLR resonator ($N$ = 3) for various $t_1 = t_3$ when $t_2 = 0.97$.

CSLR resonators with two SLRs ($N$ = 2) can be regarded as single FP cavities without mode splitting.[23,29] Here, we start from the CSLR resonators with three SLRs ($N$ = 3). Based on Eqs. (1) – (6), the calculated power transmission spectra and group delay spectra of the CSLR resonators with three SLRs ($N$ = 3) are depicted in Fig. 2. The structural parameters are chosen as follows: $L_{s1} = L_{s2} = L_{s3} = 129.66$ μm, and $L_1 = L_2 = 100$ μm. For single-mode silicon photonic nanowire waveguides with a cross-section of 500 nm × 260 nm, we use values based on our previously fabricated devices for the waveguide group index of the transverse electric (TE) mode ($n_g$ = 4.3350) and the propagation loss ($\alpha$ = 55 m$^{-1}$ (2.4 dB/cm)). The same values of $n_g$ and $\alpha$ are also used for the calculations of other transmission and group delay spectra in this section. The calculated power transmission spectra of the CSLR resonator ($N$ = 3) for various $t_2$ when $t_1 = t_3 = 0.87$ are shown in Fig. 2(a). The corresponding group delay spectra are shown in Fig. 2(b). It is clear that different degrees of mode splitting can be achieved by varying $t_2$. As $t_2$ decreases (i.e., the coupling strength increases), the spectral range between the two adjacent resonant peaks decreases until the split peaks finally merge into one. By further decreasing $t_2$, the Q factor, extinction ratio, and group delay of the combined single resonance increases, together with an increase in the insertion loss. In particular, when $t_2$ = 0.77, a band-pass Butterworth filter[34] with a flat-top filter shape can be realized, which is desirable for signal filtering in optical communications systems.[35,36] On the other hand, when $t_2$ = 0.742, the CSLR resonator exhibits a flat-top group delay spectrum, which can be used as a Bessel filter for optical buffering.[37,38] When $t_2 = \sqrt{1/2}$, SLR$_2$ works as a total reflector, and so there is null transmission for the CSLR resonator. The same goes for $t_1 = \sqrt{1/2}$ or $t_3 = \sqrt{1/2}$. Figure 2(c) shows the calculated power transmission spectra of the CSLR resonator ($N$ = 3) for various $t_1 = t_3$ when $t_2$ = 0.97. The group delay spectra are depicted in Fig. 2(d) accordingly. One can see that decreasing $t_1$ and $t_3$ (i.e., enhancing the coupling strengths) results in increased Q factor, extinction ratio, and group delay, at the expense of an increase in the insertion loss. The sharpening of the filter shape can be attributed to coherent interference within the coupled resonant cavities, which could be useful for implementation of high-Q filters.[7,32].

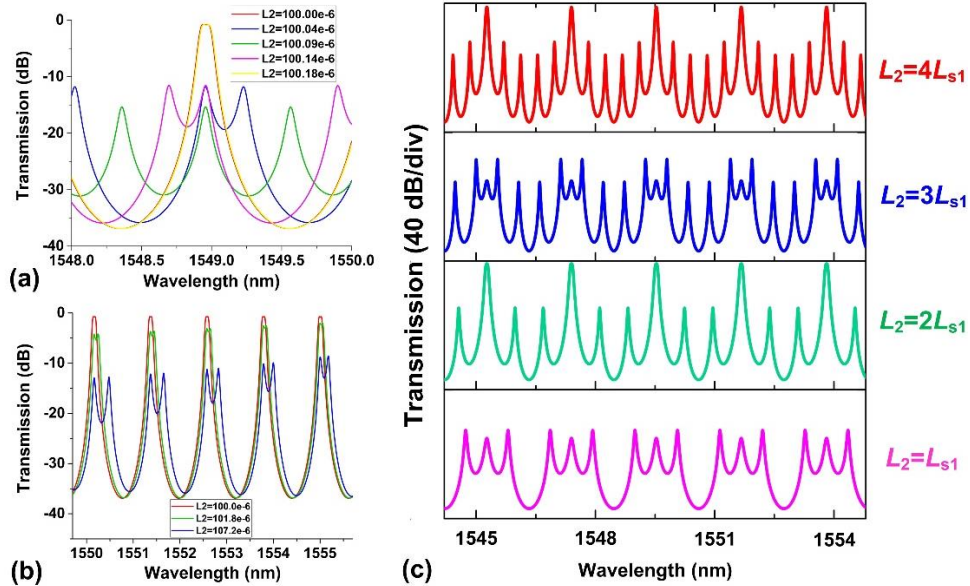

FIG. 3. (a) Calculated power transmission spectra of the CSLR resonator ($N = 3$) for various $L_2$ from 100.00 μm to 100.18 μm when $L_1 = 100.00$ μm. (b) Calculated power transmission spectra of the CSLR resonator ($N = 3$) for various $L_2$ from 100.0 μm to 107.2 μm when $L_1 = 100.0$ μm. (c) Calculated power transmission spectra of the CSLR resonator ($N = 3$) for various $L_2$ from $L_{s1}$ to $4L_{s1}$ when $L_1 = 0$ μm.

Figure 3 shows the calculated power transmission spectra of asymmetric CSLR resonators ($N = 3$) when $L_1 \neq L_2$. For comparison, we use the same field transmission coefficients $[t_1, t_2, t_3] = [0.87, 0.77, 0.87]$ in the calculation. In Fig. 3(a), we plot the calculated power transmission spectra around one resonance when there are relatively small differences between $L_1$ and $L_2$. It can be seen that the differences between $L_1$ and $L_2$ lead to different filter shapes of the CSLR resonator. In particular, when $L_2 = 100.18$ μm, the transmission spectrum of the CSLR resonator is almost the same as when $L_2 = 100.00$ μm. This is because in such a condition the difference between the phase along $L_1$ and $L_2$ is approximately π. Considering that the physical cavity length is half of the effective cavity length for a SW resonator,[23] the effective phase difference is about 2π, and so there are almost the same transmission spectra resulting from coherent interference within in the resonant cavity. The calculated power transmission spectra in Fig. 3(a) also indicate that the filter shape of the CSLR resonator can be tuned or optimized by introducing thermo-optic micro-heaters[19,33] or carrier-injection electrodes[39,40] along $L_{1,2}$ to tune the phase shift. Figure 3(b) presents the calculated power transmission spectra when there are relatively large differences between $L_1$ and $L_2$. Due to the Vernier-like effect between the FPC$_1$ and FPC$_2$, diverse mode splitting filter shapes are achieved at different resonances of the transmission spectra, which can be utilized to select resonances with desired filter shapes for passive photonic devices.[41] Such differences in the filter shapes become more obvious for an increased difference between $L_1$ and $L_2$. In Fig. 3(c), we plot the calculated power transmission spectra when $L_1 = 0$ and $L_2 = mL_{s1}$ ($m = 1, 2, 3, 4$). Since the effective cavity length of FPC$_i$ equals to $L_{si} + 2L_i + L_{si+1}$ ($i = 1, 2$),[23] the various sets of $L_1$ and $L_2$ in Fig. 3(c) correspond to the conditions that the effective cavity length of FPC$_2$ is integer multiples of that of FPC$_1$. We can see that there are split resonances with different numbers of transmission peaks in the spectra. Unlike the split resonances in Fig. 3(b), there are identical filter shapes in each period.

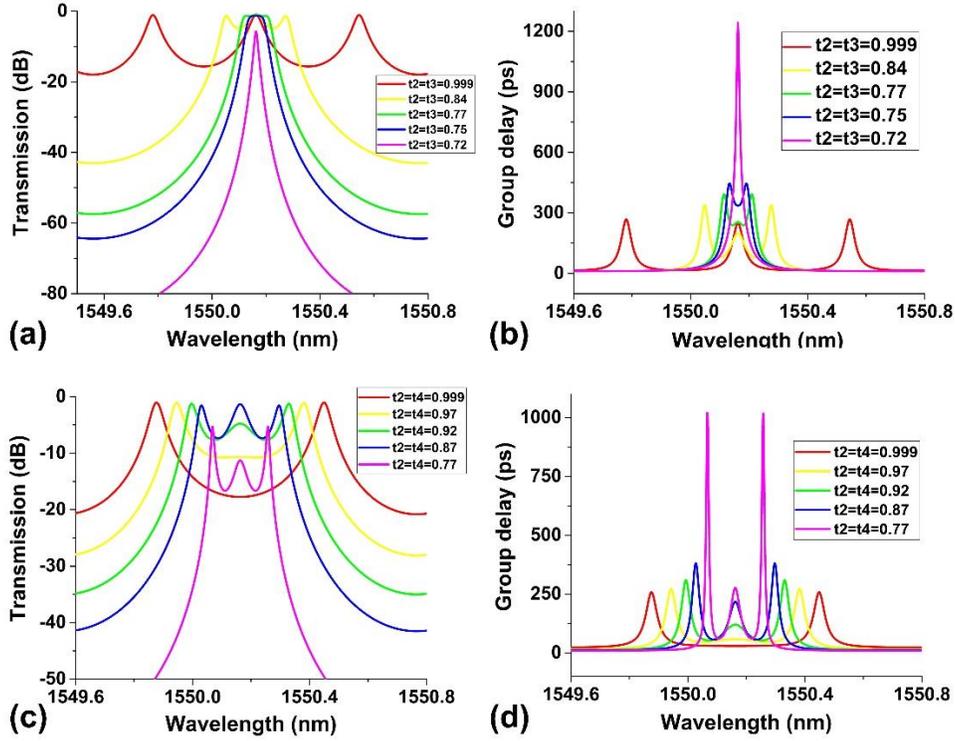

FIG. 4. (a) Calculated power transmission spectra of the CSLR resonator ($N = 4$) for various $t_2 = t_3$ when $t_1 = t_4 = 0.87$. (b) Calculated group delay spectra of the CSLR resonator ($N = 4$) for various $t_2 = t_3$ when $t_1 = t_4 = 0.87$. (c) Calculated power transmission spectra of the CSLR resonator ($N = 4$) for various $t_2 = t_4$ when $t_1 = t_3 = 0.85$. (d) Calculated group delay spectra of the CSLR resonator ($N = 4$) for various $t_2 = t_4$ when $t_1 = t_3 = 0.85$.

Figure 4 shows the calculated power transmission spectra of the CSLR resonators with four SLRs ($N = 4$). In the calculation, we assume $L_{s1} = L_{s2} = L_{s3} = L_{s4} = 129.66$ μm, and $L_1 = L_2 = L_3 = 100$ μm. To simplify the comparison, we only show plots for the conditions that $t_1 = t_4$, $t_2 = t_3$ (Figs. 4(a) and (b)) and $t_1 = t_3$, $t_2 = t_4$ (Figs. 4(c) and (d)). The calculated power transmission spectra for various $t_2 = t_3$ when $t_1 = t_4 = 0.87$ are shown in Fig. 4(a). The corresponding group delay spectra are provided in Fig. 4(b). It can be seen that the three split resonant peaks gradually merge to a single one as $t_2$ and $t_3$ decrease (i.e., the coupling strengths increase). After that, by further decreasing $t_2$ and $t_3$, the Q factor, extinction ratio, and group delay of the combined single resonance increase, together with an increase in the insertion loss. This trend is similar to that in Figs. 3(a) and (b) for $N = 3$. Figure 4(c) depicts the calculated power transmission spectra for various $t_2 = t_4$ when $t_1 = t_3 = 0.85$. The calculated group delay spectra are shown in Fig. 4(d) accordingly. For this condition, the CSLR resonator is no longer axisymmetric, and so the trend in Figs. 4(c) and (d) is different from that in Figs. 4(a) and (b). As $t_2$ and $t_4$ decrease (i.e., the coupling strengths increase), the transmission peak in the centre starts to appear and then becomes more pronounced, together with a decreased spectral range between the resonant peaks on both sides.

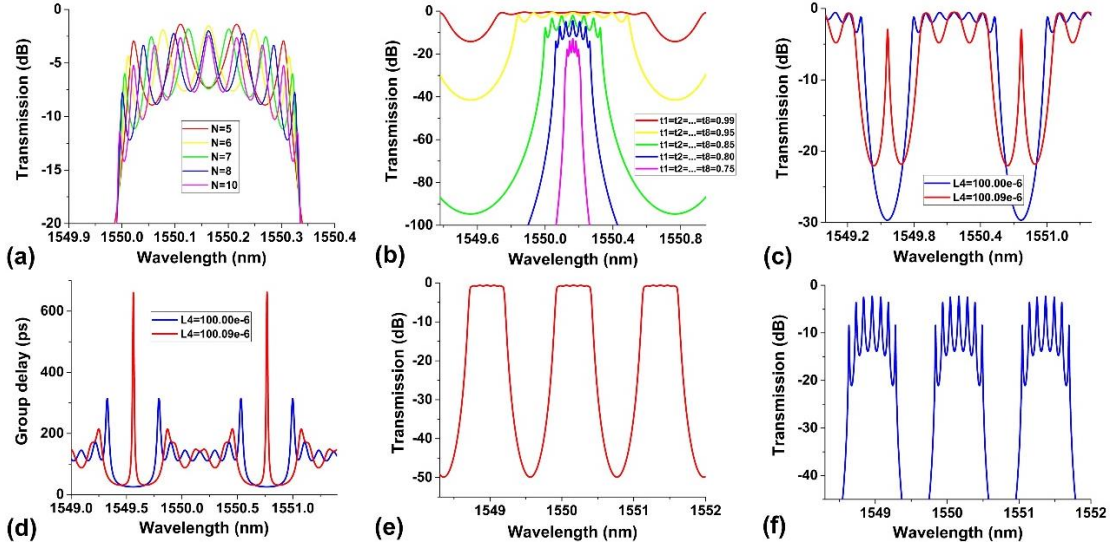

FIG. 5. (a) Calculated power transmission spectra of the CSLR resonator for various $N$ when $t_1 = t_2 =...= t_N = 0.85$. (c) Calculated power transmission spectra of the CSLR resonator ($N = 8$) for different $t_1 = t_2 =...= t_8$. (c) Calculated power transmission spectra of the CSLR resonator ($N = 8$) for enhanced light trapping. (d) Calculated group delay spectra of the CSLR resonator ($N = 8$) in (c). (e) Calculated power transmission spectra of 8th-order Butterworth filter based on the CSLR resonator ($N = 8$). (f) Calculated power transmission spectra of the CSLR resonator ($N = 8$) with multiple transmission peaks.

Figure 5(a) shows the calculated power transmission spectra of the CSLR resonators with different numbers of SLRs ($N$). In the calculation, we use SLRs and connecting waveguides with the same lengths as those in Figs. 2 and 4. We also assume that $t_1 = t_2 =...= t_N = 0.85$. It can be seen that as $N$ increases, the number of split resonances within one FSR also increases. For a CSLR resonator consisting of $N$ SLRs, the maximum number of split resonances within one FSR is $N$-1. The differences between the maximum transmission of different resonant peaks are determined by the waveguide propagation loss, and these can be mitigated by decreasing the waveguide propagation loss ($\alpha$) to the point where, in the limit of zero loss, they would no longer exist. In Fig. 5(b), we plot the calculated power transmission spectra of the CSLR resonator ($N = 8$) for different $t_1 = t_2 =...= t_8$. As $t_i$ ($i = 1, 2, ..., 8$) increases (i.e., the coupling strengths decrease), the bandwidth of the passband also increases, together with a decrease in insertion loss. In principle, the bandwidth of the passband is limited by the FSR of the CSLR resonator. Figures 5(c) − (f) show three specific optical filters designed based on the CSLR resonators with eight SLRs ($N = 8$). The filter in Figs. 5(c) and (d) is designed for enhanced light trapping by introducing an additional π/2 phase shift along the centre FPC (i.e., $L_4$ for $N = 8$), which is similar to enhancing light trapping in photonic crystals by introducing defects.[17] In the calculation, we assume that $t_1 = t_2 =...= t_8 = 0.97$. With enhanced light trapping, there are increased time delays and enhanced light-matter interactions, which are useful in nonlinear optics and laser excitation.[10,11,27,28] In Fig. 5(c), one can see that there are central transmission peaks induced by an additional phase shift along $L_4$, which correspond to a group delay 2.1 times higher than that of the CSLR resonator without the additional phase shift in Fig. 5(d). This group delay can be increased further by using more cascaded SLRs. The filter in Fig. 5(e) is an 8th-order Butterworth filter with a flat-top filter shape. The field transmission coefficients of SLR$_i$ ($i = 1, 2, ..., 8$) are [$t_1, t_2, t_3, t_4, t_5, t_6, t_7, t_8$] = [0.98, 0.94, 0.91, 0.90, 0.90, 0.91, 0.94, 0.98], respectively. Figure 5(f) shows the designed optical filter with multiple transmission peaks in the spectrum. Each transmission peak has a high extinction ratio of over 10 dB. The field transmission coefficients of $SLR_i$ ($i$ = 1, 2, …, 8) in this filter are [$t_1, t_2, t_3, t_4, t_5, t_6, t_7, t_8$] = [0.84, 0.935, 0.945, 0.955, 0.955, 0.945, 0.935, 0.84], respectively. By tailoring the transmission of individual resonant peaks via changing $t_i$ and $\varphi_i$, these CSLR resonators could potentially find applications in RF spectral shaping and broadband arbitrary RF waveform generation[42,43]. Based on two-photon absorption (TPA)-induced free carrier dispersion (FCD) [44], CSLR resonators with multiple transmission peaks could also be used for wavelength multicasting in wavelength division multiplexing (WDM) systems.[9,45] It is also worth mentioning that the narrow bandwidth between the split resonances arises from coherent interference within the CSLR resonators. For ring resonators, such a narrow bandwidth can only be achieved by using much larger loop circumferences, thus leading to much larger device footprints. In addition, the CSLR resonator is a SW resonator, and so the cavity length is nearly half that of a TW resonator (e.g., ring resonator) with the same FSR, which enables even more compact device footprints.

## III. DEVICE FABRICATION

We fabricated a series of CSLR filters based on the designs in Section II, on an SOI wafer with a 260-nm-thick top silicon layer and 3-μm-thick buried oxide layer. The device fabrication involved standard complementary metal-oxide-semiconductor (CMOS) processes only, with the exception that the device pattern was defined using electron-beam lithography (EBL). The micrographs for the fabricated devices with four and eight SLRs are shown in Figs. 6(a) and (b), respectively. A zoom-in micrograph for the SLR is shown in Fig. 6(c). In our fabrication, EBL (Vistec EBPG 5200) was employed to define the device pattern on positive photoresist (ZEP520A), followed by a reactive ion etching (RIE) process to transfer the device layout to the top silicon layer of the SOI wafer. Grating couplers for TE polarization were employed at the ends of input and output ports to couple light into and out of the chip with single mode fibres, respectively. The grating couplers were fabricated by another EBL step, together with a second RIE process. Gold markers, prepared by metal lift-off after photolithography and electron beam evaporation, were employed for alignment between the two times EBL. A 1.5-μm-thick silica layer is deposited by plasma enhanced chemical vapor deposition (PECVD) to cover the whole device as upper cladding. For all the devices, the width of the waveguides is 500 nm and the gap size of all the directional couplers was 100 nm. Different coupling strengths for different SLRs were achieved by changing $L_{ci}$ ($i$ = 1, 2, …, $N$) in Fig. 1. Compared with silicon waveguide Bragg gratings based on small physical corrugations, the fabrication of self-coupled nanowire waveguides doesn't require high lithography resolution and shows relatively high tolerance to fabrication imperfections such as lithographic smoothing effects and quantization errors due to the finite grid size.[30] For the calculations in Section II, $L_{ci}$ ($i$ = 1, 2, …, $N$) are included in $L_i$ ($i$ = 1, 2, …, $N$-1). Their relation is given by:

$$L_i = L_i' + L_{ci} + L_{ci+1}, \quad i = 1, 2, .., N\text{-}1 \tag{7}$$

where $L_i'$ ($i$ = 1, 2, …, $N$-1) are the lengths of the connecting waveguides excluding the straight coupling lengths. For practical fabrication, we used the same Sagnac loop structure for each SLR, with the differences in $L_{ci}$ ($i$ = 1, 2, …, $N$) compensated by slightly changing $L_i'$ ($i$ = 1, 2, …, $N$-1) according to Eq. (7).

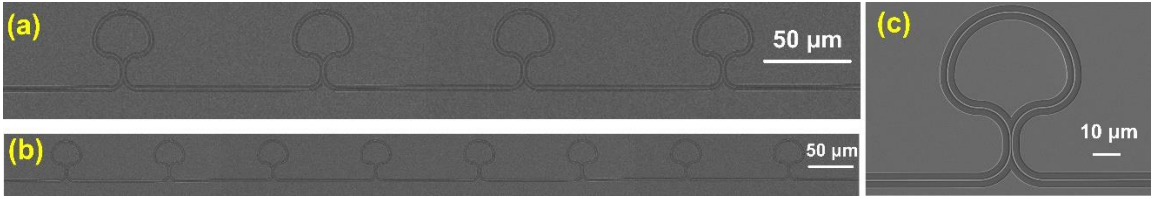

FIG. 6. (a) Micrograph for one of the fabricated CSLR resonators with four SLRs. (b) Micrograph for one of the fabricated CSLR resonators with eight SLRs. (c) Zoom-in micrograph for the SLR.

## IV. DEVICE CHARACTERISATION

The normalized transmission spectra for the fabricated CSLR resonators with three SLRs ($N = 3$) are shown in Figs. 7(a) and (b) by the blue solid curves. The spectra were measured by sweeping a fast-scan continuous-wave (CW) laser (Keysight 81608A) with a power of ~0 dBm. The output powers from the devices under test were recorded using a high-sensitivity optical power meter (Keysight N7744A). The measured transmission spectra are normalized and plot after excluding the losses caused by the grating couplers, which was ~6 dB each, or ~12 dB for both. The normalized spectra are then fit by the red dashed curves calculated based on Eqs. (1) – (6). The fit parameters are listed in Table II, which are close to our expectations from the design. For $n_g$, $\alpha$, and $t_i$, the difference between the fit and designed values are smaller than 0.01, 20 m$^{-1}$ (0.6 dB/cm), and 0.05, respectively. The residual differences between the fit $n_g$ and $\alpha$ can be attributed to slight variations between the fabricated samples. In Fig. 7(a), various mode splitting spectra of the fabricated devices with different $L_{c2}$ are obtained, which are consistent with the theory in Fig. 2(a). The measured spectra of the fabricated devices with different $L_{c1} = L_{c3}$ in Fig. 7(b) also agree well with the theory in Fig. 2(c). These experimental results verify that the transmission spectra of the CSLR resonators can be tailored by changing the coupling strengths of the directional couplers in the SLRs. Since we have demonstrated in Ref. 29 that dynamic tuning of the coupling strengths can be realized by using interferometric couplers to replace the directional couplers and tuning them in a differential mode, tuning of the transmission spectra of the CSLR resonators can also be achieved in the same way.

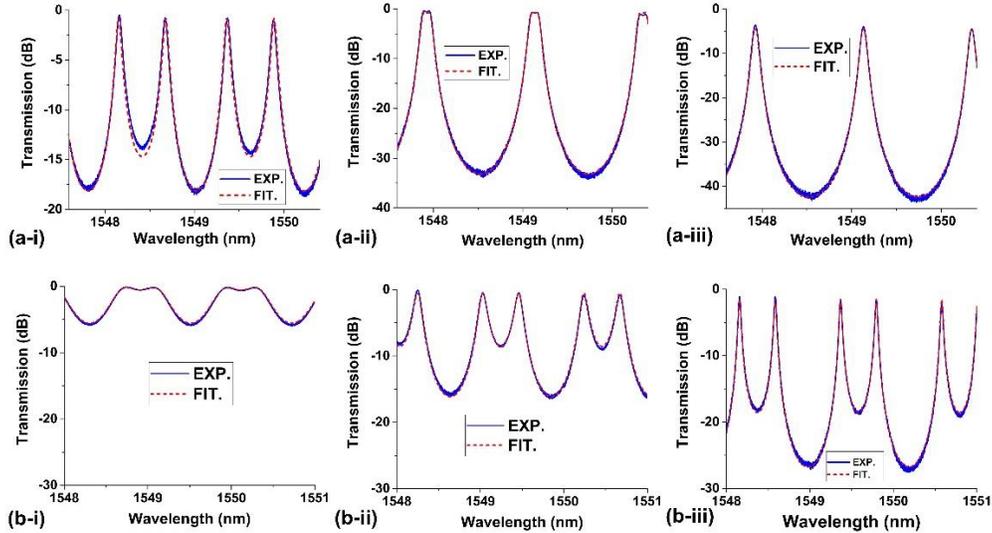

FIG. 7. (a) Measured (solid curve) and fit (dashed curve) transmission spectra of three fabricated CSLR resonators ($N = 3$) with different $L_{c2}$. (b) Measured (solid curve) and fit (dashed curve) transmission spectra of three fabricated CSLR resonators ($N = 3$) with different $L_{c1} = L_{c3}$.

TABLE II. Fit parameters of the measured transmission spectra in Figs. 7, 8, and 9

| Figure | Fit parameters | |
|---|---|---|
| Fig. 7(a) | (i) | $n_g$ = 4.3202, $\alpha$ = 62 m$^{-1}$ (~2.7 dB/cm), [$t_1$, $t_2$, $t_3$] = [0.876, 0.995, 0.876] |
| | (ii) | $n_g$ = 4.3221, $\alpha$ = 64 m$^{-1}$ (~2.8 dB/cm), [$t_1$, $t_2$, $t_3$] = [0.883, 0.785, 0.883] |
| | (iii) | $n_g$ = 4.3220, $\alpha$ = 64 m$^{-1}$ (~2.8 dB/cm), [$t_1$, $t_2$, $t_3$] = [0.886, 0.734, 0.886] |
| Fig. 7(b) | (i) | $n_g$ = 4.3178, $\alpha$ = 59 m$^{-1}$ (~2.6 dB/cm), [$t_1$, $t_2$, $t_3$] = [0.982, 0.975, 0.982] |
| | (ii) | $n_g$ = 4.3180, $\alpha$ = 59 m$^{-1}$ (~2.6 dB/cm), [$t_1$, $t_2$, $t_3$] = [0.902, 0.976, 0.902] |
| | (iii) | $n_g$ = 4.3180, $\alpha$ = 60 m$^{-1}$ (~2.6 dB/cm), [$t_1$, $t_2$, $t_3$] = [0.829, 0.974, 0.829] |
| Fig. 8(a) | $L_2$ = 100.0 µm: $n_g$ = 4.3252, $\alpha$ = 58 m$^{-1}$ (~2.6 dB/cm), [$t_1$, $t_2$, $t_3$] = [0.882, 0.788, 0.882] | |
| | $L_2$ = 107.2 µm: $n_g$ = 4.3249, $\alpha$ = 58 m$^{-1}$ (~2.6 dB/cm), [$t_1$, $t_2$, $t_3$] = [0.880, 0.789, 0.880] | |
| Fig. 8(b) | $L_2$ = 259.33 µm, $n_g$ = 4.3252, $\alpha$ = 64 m$^{-1}$ (~2.8 dB/cm), [$t_1$, $t_2$, $t_3$] = [0.882, 0.783, 0.882] | |
| Fig. 9(a) | (i) | $n_g$ = 4.3278, $\alpha$ = 60 m$^{-1}$ (~2.6 dB/cm), [$t_1$, $t_2$, $t_3$, $t_4$] = [0.882, 0.992, 0.992, 0.882] |
| | (ii) | $n_g$ = 4.3280, $\alpha$ = 58 m$^{-1}$ (~2.6 dB/cm), [$t_1$, $t_2$, $t_3$, $t_4$] = [0.931, 0.832, 0.832, 0.931] |
| | (iii) | $n_g$ = 4.3272, $\alpha$ = 60 m$^{-1}$ (~2.6 dB/cm), [$t_1$, $t_2$, $t_3$, $t_4$] = [0.834, 0.958, 0.958, 0.834] |
| Fig. 9(b) | $n_g$ = 4.3300, $\alpha$ = 65 m$^{-1}$ (~2.8 dB/cm), $L_4$ = 100.09 µm, | |
| | [$t_1$, $t_2$, $t_3$, $t_4$, $t_5$, $t_6$, $t_7$, $t_8$] = [0.962, 0.962, 0.962, 0.962, 0.962, 0.962, 0.962, 0.962] | |
| Fig. 9(c) | $n_g$ = 4.3288, $\alpha$ = 68 m$^{-1}$ (~3.0 dB/cm), | |
| | [$t_1$, $t_2$, $t_3$, $t_4$, $t_5$, $t_6$, $t_7$, $t_8$] = [0.998, 0.967, 0.942, 0.913, 0.913, 0.942, 0.967, 0.998] | |
| Fig. 9(d) | $n_g$ = 4.3299, $\alpha$ = 64 m$^{-1}$ (~2.8 dB/cm), | |
| | [$t_1$, $t_2$, $t_3$, $t_4$, $t_5$, $t_6$, $t_7$, $t_8$] = [0.874, 0.941, 0.972, 0.982, 0.982, 0.972, 0.941, 0.874] | |

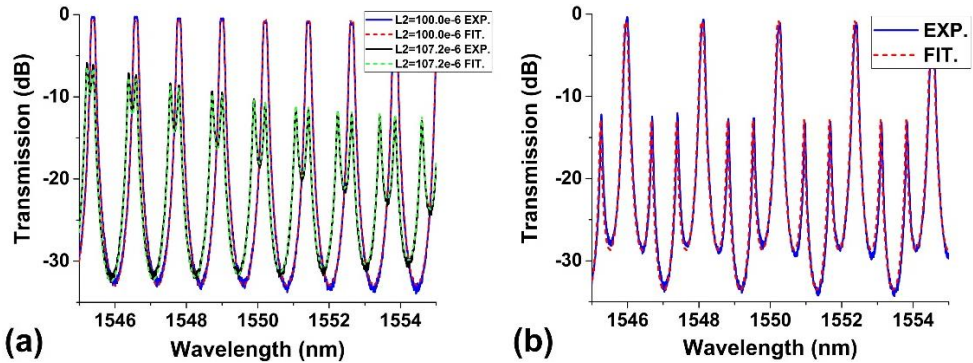

FIG. 8. (a) Measured (solid curve) and fit (dashed curve) transmission spectra of two fabricated CSLR resonators ($N$ = 3) with different $L_2$ = 100.0 µm and 107.2 µm. (b) Measured (solid curve) and fit (dashed curve) transmission spectra of a fabricated CSLR resonator ($N$ = 3) with $L_1{'}$ = 0 µm and $L_2$ = 295.33 µm.

The measured and fit transmission spectra of the fabricated asymmetric CSLR resonator ($N$ = 3) with $L_1$ =100.0 µm and $L_2$ = 107.2 µm are shown in Fig. 8(a). For comparison, the measured and fit transmission spectra of a symmetric CSLR resonator ($N$ = 3) with $L_1$ = $L_2$ =100.0 µm are also shown. The fit parameters are also given in Table II. For the symmetric CSLR resonator, there is negligible difference between the filter shapes of different resonances. In contrast, due to the Vernier-like effect, there are obvious differences between the filter shapes of different resonances for the fabricated asymmetric CSLR resonator. These experimental results match with the theory in Fig. 3(b). Figure 8(b) shows the measured and fit transmission spectra of the fabricated asymmetric CSLR resonator ($N$ = 3) with $L_1{'}$= $L_1$-$L_{c1}$-$L_{c2}$ = 0 µm and $L_2$ = 295.33 μm. In the device pattern, the loop region of $SLR_1$ was rotated to the bottom side of $SLR_2$. It can be seen that the measured spectrum shows good agreement with the theory in Fig. 3(c), with the discrepancies mainly arising from the grating coupler spectral response as well as slight variations in coupling coefficients with wavelength.[46] The dispersion of the SOI nanowire waveguides is another factor that could account for the discrepancies, since we used $n_g$ rather than the wavelength-dependent effective index to match the FSR in our theoretical calculations.

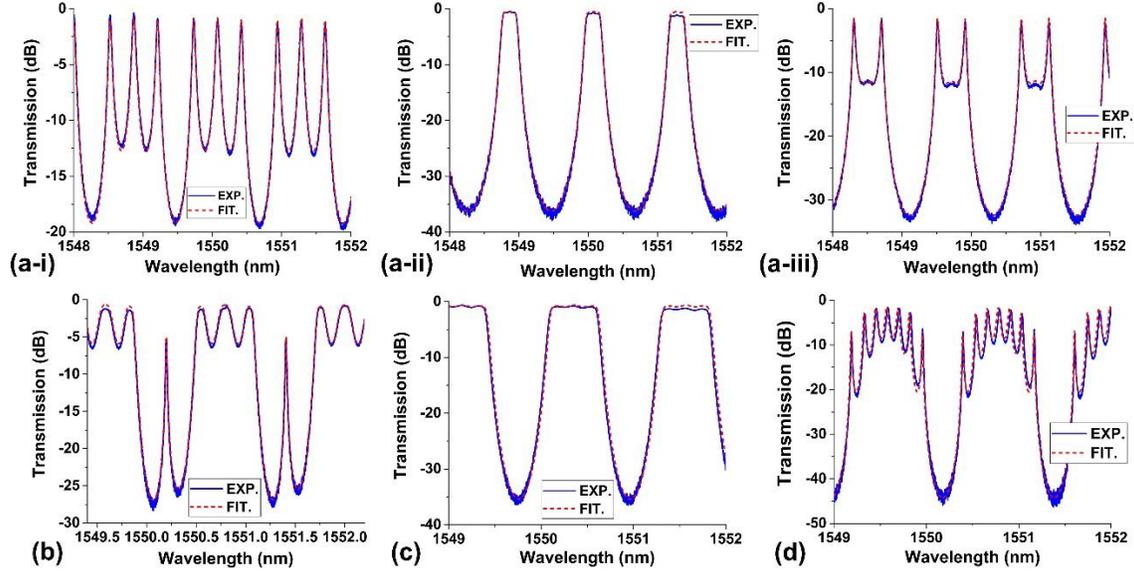

FIG. 9. (a) Measured (solid curve) and fit (dashed curve) transmission spectra of three fabricated CSLR resonators ($N = 4$) with different $L_{ci}$ ($i = 1, 2, 3, 4$). (b) Measured (solid curve) and fit (dashed curve) transmission spectra of a fabricated CSLR resonator ($N = 8$) for enhanced light trapping. (c) Measured (solid curve) and fit (dashed curve) transmission spectra of a fabricated CSLR resonator ($N = 8$) with flat-top filter shape. (d) Measured (solid curve) and fit (dashed curve) transmission spectra of a fabricated CSLR resonator ($N = 8$) with multiple split resonances.

The measured and fit transmission spectra of the fabricated CSLR resonators with four SLRs ($N = 4$) are shown in Fig. 9(a). The fitting parameters are listed in Table II. One can see that diverse filter shapes are obtained for the fabricated devices with different coupling lengths, and all the measured spectra agree well with the theory in Fig. 4. Figures 9(b) − (d) show the measured and fit transmission spectra of the fabricated CSLR resonators with eight SLRs ($N = 8$). The fitting parameters are also provided in Table II. The device in Fig. 8(b) was designed for enhanced light trapping, and the measured transmission spectrum is similar to the calculated spectrum in Fig. 5(c). The measured filter shape in Fig. 9(b) exhibits a slight asymmetry and this is because the additional phase shift along $L_4$ is not exactly $π/2$. By introducing thermo-optic micro-heaters or carrier-injection electrodes along $L_4$ to tune the phase shift, the symmetry of the filter shape can be improved further. The device in Fig. 9(c) was designed to perform as an 8th-order Butterworth filter with a flat-top filter shape. As can be seen, the passband is almost flat, which is close to the calculated spectrum in Fig. 5(e). The 3-dB bandwidth is ~0.7 nm, which is significantly narrower than what can typically be achieved by silicon waveguide Bragg gratings[30]. By either increasing the relevant $t_i$ or by increasing the number of SLRs, the 3-dB bandwidth can be further improved. The slight unevenness of the top of the transmission band can be attributed to discrepancies between the designed and practical coupling coefficients. Figure 9(d) shows the measured transmission spectrum with multiple resonant peaks. The minimum extinction ratio of the transmission peaks is ~7.8 dB, which is slightly lower than that in Fig. 5(f). This is mainly because the waveguide propagation loss of the fabricated devices ($\alpha = 64$ m$^{-1}$) is slightly higher than we assumed in the calculation ($\alpha = 55$ m$^{-1}$). By further reducing the propagation loss, higher extinction ratios of the split resonances can be obtained.

## V. CONCLUSION

In summary, we design and fabricate sophisticated and high performance optical filters in SOI nanowires through the use of mode splitting in integrated CSLR resonators, by designing the reflectivity of the SLRs and the phase shifts along the connecting waveguides. These filters are extremely useful for a wide range of applications including enhanced light trapping, flat-top filtering, Q factor enhancement, and signal reshaping. We theoretically analyse and practically fabricate devices with up to eight SLRs. We measure the transmission spectra of the fabricated devices and obtain versatile filter shapes corresponding to diverse mode splitting conditions. The experimental results are consistent with theory and validate the CSLR resonator as a powerful and versatile approach to realise multi-functional SW filters for flexible spectral engineering in photonic integrated circuits.

## ACKNOWLEDGMENTS

This work was supported by the Australian Research Council Discovery Projects Program (DP150104327). We also acknowledge the Melbourne Centre for Nanofabrication (MCN) for the support in device fabrication and the Swinburne Nano Lab for the support in device characterisation.

[1]W. Bogaerts, P. De Heyn, T. Van Vaerenbergh, K. De Vos, S. Selvaraja, T. Claes, P. Dumon, P. Bienstman, D. Van Thourhout, and R. Baets, Laser Photon. Rev. **6**, 47 (2012).
[2]S. Feng, T. Lei, H. Chen, H. Cai, X. Luo, and A. W. Poon, Laser Photon. Rev. **6**, 145 (2012).
[3]Q. Li, T. Wang, Y. Su, M. Yan, and M. Qiu, Opt. Exp. **18**, 8367 (2010).
[4]B. Peng, ŞK Özdemir, W. Chen, F. Nori, and L. Yang, Nat. Commun. **5**, 1 (2014).
[5]M. Limonov, M. Rybin, A. Poddubny, and Y. Kivshar, Nat. Photon. **11**, 543 (2017).
[6]M. Fleischhauer, A. Imamoglu, J. P. Marangos, Rev. Mod. Phys. **77**, 633 (2005).
[7]L. Barea, F. Vallini, G. de Rezende, and N. Frateschi, Photon. J. **5**, 2202717 (2013).
[8]L. Barea, F. Vallini, P. Jarschel, and N. Frateschi, App. Phys. Lett. **103**, 201102 (2013).
[9]M. Souza, L. Barea, F. Vallini, G. Rezende, G. Wiederhecker, and N. Frateschi, Opt. Exp. **22**, 10430 (2014).
[10] M. D. Lukin and A. Imamoğlu, Nature **413**, 273 (2001).
[11] C. Monat, C. Grillet, B. Corcoran, D. J. Moss, B. J Eggleton, T. P. White, and T. F. Krauss, Opt. Exp. **18**, 6831 (2010).
[12]X. Xue, Y. Xuan, P. Wang, Y. Liu, D. Leaird, M. Qi, and A. Weiner, Laser Photon. Rev. **9**, 123 (2015).
[13]S. Kim, K. Han, C. Wang, J. Villegas, X. Xue, C. Bao, Y. Xuan, D. Leaird, A. Weiner, and M. Qi, Nat. Commun. **8**, 1 (2017).
[14]M. Souza, G. Rezende, L. Barea, G. Wiederhecker, and N. Frateschi, Opt. Exp. **24**, 18960 (2016).
[15]L. Zhang, M. Song, T. Wu, L. Zou, R. Beausoleil, and A. Willner, Opt. Lett. **33**, 1978 (2008).
[16]L. Zhou, T. Ye, and J. Chen, Opt. Lett. **36**, 13 (2011).
[17]J. Song, L. Luo, X. Luo, H. Zhou, X. Tu, L. Jia, Q. Fang, and G. Lo, Opt. Exp. **22**, 24202 (2014).
[18]J. Wu, B. Liu, J. Peng, J. Mao, X. Jiang, C. Qiu, C. Tremblay, and Y. Su, J. Lightw. Technol. **33**, 3542 (2015).
[19]M. Souza, G. Rezende, L. Barea, A. Zuben, G. Wiederhecker, and N. Frateschi, Opt. Lett. **40**, 3332 (2015).
[20]A. Li and W. Bogaerts, Opt. Exp. **25**, 31688 (2017).
[21]A. Li and W. Bogaerts, APL Photonics **2**, 096101 (2017).
[22]S. Zheng, Z. Ruan, S. Gao, Y. Long, S. Li, M. He, N. Zhou, J. Du, L. Shen, X. Cai and J. Wang, Opt. Exp. **25**, 25655 (2017).
[23]X. Sun, L. Zhou, J. Xie, Z. Zou, L. Lu, H. Zhu, X. Li, and J. Chen, Opt. Lett. **38**, 567 (2013).
[24]Y. Ni, App. Opt. **19**, 3425 (1980).
[25]V. Van, *Optical Microring Resonators: Theory, Techniques, and Applications*, CRC Press, 2016), p.53.
[26]Z. Wang, B. Tian, M. Pantouvaki, W. Guo, P. Absil, J. Van Campenhout, C. Merckling, and D. Van Thourhout, Nat. Photon. **9**, 837 (2015).
[27]S. Wu, S. Buckley, J. Schaibley, L. Feng, J. Yan, D. Mandrus, F. Hatami, W. Yao, J. Vučković, A. Majumdar, and X. Xu, Nature **520**, 69 (2015).
[28]T. Gu, N. Petrone, J. McMillan, A. van der Zande, M. Yu, G. Lo, D. Kwong, J. Hone, and C. Wong, Nat. Photon. **6**, 554 (2012).
[29]X. Jiang, J. Wu, Y. Yang, T. Pan, J. Mao, B. Liu, R. Liu, Y. Zhang, C. Qiu, C. Tremblay, and Y. Su, Opt. Exp. **24**, 2183 (2016).

[30] X. Wang, Y. Wang, J. Flueckiger, R. Bojko, A. Liu, A. Reid, J. Pond, N. Jaeger, and L. Chrostowski, Opt. Lett. **39**, 5519 (2014).
[31] L. Lu, F. Li, M. Xu, T. Wang, J. Wu, L. Zhou, and Y. Su, Photon. Technol. Lett. **24**, 1765 (2012).
[32] J. Wu, T. Moein, X. Xu, G. Ren, A. Mitchell, and D. J. Moss, APL Photonics **2**, 056103 (2017)
[33] J. Wu, P. Cao, T. Pan, Y. Yang, C. Qiu, C. Tremblay, and Y. Su, Photon. Res. **3**, 9 (2015).
[34] H. Liu and A. Yariv, Opt. Exp. **19**, 17653 (2011).
[35] B. Little, S. Chu, P. Absil, J. Hryniewicz, F. Johnson, F. Seiferth, D. Gill, V.Van, O. King, and M. Trakalo, Photon. Technol. Lett. **16**, 2263 (2004).
[36] F. Xia, M. Rooks, L. Sekaric, and Y. Vlasov, Opt. Exp. **15**, 11934 (2007).
[37] F. Xia, L. Sekaric, Y. Vlasov, Nat. Photon. **1**, 65 (2007).
[38] J. Poon, J. Scheuer, Y. Xu, and A. Yariv, J. Opt. Soc. Am. B, **21**, 1665 (2004).
[39] L. Zhou and A. W. Poon, Opt. Exp. **15**, 9194 (2007).
[40] J. Wu, P. Cao, X. Hu, X. Jiang, T. Pan, Y. Yang, C. Qiu, C. Tremblay, Y. Su, Opt. Exp. **22**, 26254 (2014).
[41] J. Wu, P. Cao, X. Hu, T. Wang, M. Xu, X. Jiang, F. Li, L. Zhou, and Y. Su, Photon. Technol. Lett. **25**, 580 (2013).
[42] M. Khan, H. Shen, Y. Xuan, L. Zhao, S. Xiao, D. Leaird, A. Weiner, and M. Qi, Nat. Photon. **4**, 117 (2010).
[43] J.Wang, H. Shen, L. Fan, R. Wu, B. Niu, L. Varghese, Y. Xuan, D. Leaird, X. Wang, F. Gan, A. Weiner, and M. Qi, Nat. Commun. **6**, 1 (2015).
[44] J. Wu, X. Jiang, T. Pan, P. Cao, L. Zhang, X. Hu, Y. Su, Chi. Sci. Bull. **59**, 2702 (2014).
[45] Q. Li, Z. Zhang, F. Liu, M. Qiu, and Y. Su, App. Phys. Lett. **93**, 081113 (2008).
[46] Q. Chen, Y. Yang, and Y. Huang, App. Phys. Lett. **89**, 061118 (2006).